\documentclass[aip,graphicx]{revtex4-1}

\usepackage{amsmath}
\usepackage{multirow}

\usepackage{graphicx}
\usepackage{color}

\begin{document}

\newcommand{\red}[1]{{\color{red} #1}}
\newcommand{\blue}[1]{{\color{blue} #1}}

\newcommand{\Basis}[1]{[\rm \operatorname{#1}]} 
\newcommand{\BasisKL}[2]{[{\rm \operatorname{#1} + k}{{#2}}]} 
\newcommand{\BasisKLIn}[2]{[{\rm \operatorname{#1} + k}L, {#2}]} 


\title{Using Koopmans' theorem for constructing basis sets: Approaching high Rydberg excited states of lithium with a compact Gaussian basis} 



\author{Jan Šmydke}
\email[]{jan.smydke@gmail.com}
\affiliation{Department of Radiation and Chemical Physics, Institute of Physics CAS,
  Na Slovance 1999/2, 18221 Praha 8, Czech Republic}


\date{\today}

\begin{abstract}
  For accurate {\it ab initio} description of Rydberg excited states
  the study suggests generating the appropriate diffuse basis
  functions by a cheap variational optimization of virtual orbitals of
  the corresponding ion core. By following this approach, dozens of
  converged correlated lithium Rydberg states, namely all the states
  up to 24S, 25P, 14D, 16F and 16G, not yet achieved by other {\it ab
    initio} approaches, could be obtained at the EOM-CCSD level of
  theory with compact and mostly state-selective contracted Gaussian
  basis sets. Despite its small size and Gaussian character, the
  optimized basis leads to highly accurate excitation energies that
  differ merely in the order of meV from the reference
  state-of-the-art explicitly correlated Gaussian method and even
  surpass Full-CI results on Slater basis by an order of magnitude.
\end{abstract}


\maketitle 

\section{Introduction}
\label{Introduction}


%
%

%

Investigation of Rydberg excited states of atomic and molecular
systems is mostly the domain of the quantum defect theory
(QDT),\cite{Seaton_1983} in which a motion of the highly excited
electron is modeled by an effective one-electron approach.  It is
assumed the average distance between the highly excited electron and
the positively charged ion core is sufficiently large that the
electron experiences a Coulombic field of the ion core analogous to
the field in the hydrogen atom, though effectively shielded by the
other electrons.  The energy spectrum of such a model necessarily
differs from the spectrum of hydrogen, introducing characteristic
quantum defects $\delta$ of the principal quantum number $n$ in the
energy level formula
\begin{equation}
  \label{EqnRyEnergy}
  E_{n} = - \frac{Ry}{(n - \delta)^2}
\end{equation}
where $Ry$ stands for the Rydberg constant.
By providing accurate quantum defects of a system ({\em e.g.} from
experimental transitions), the asymptotic wave functions can be
expressed analytically and the properties further evaluated. Although
the QDT proved to be robust and capable, it represents an one-electron
approach, not an {\em ab initio} theory and depends on the externally
supplied energy spectrum.  Its applicability is also limited to rather
highly excited states, in which the true Rydberg character of the
system is sufficiently retained.

By contrast, the {\em ab initio} theories for excited states either
suffer from overall poor accuracy of the computed energy spectrum or
they provide only the few lowest excited states in high precision.
This stems from the inherent complexity of the {\em ab initio}
methods, which deal with many-body systems and balance between
scalability and an accurate description of the electronic correlation.

The sharp contrast between the ability of the advanced quantum
chemical methods to describe highly correlated ground electronic
states even in difficult electronic structures on the one hand and their
failure to reliably describe the higher excited states on the other
hand grows out of the insufficiency of the commonly used basis sets to
reach and mimic the diffuse and structurally more complicated excited
state wave functions. The standard Gaussian basis sets (GTO) tend to
be extensively optimized for the ground state to describe enough
electronic correlation while keeping the number of basis functions
low.
The exponential-type basis functions (ETO) like the
Slater-type-orbitals (STO) and Coulomb-Sturmians (CS) are, of course,
of higher quality than the Gaussians due to the correct cusp at nuclei
and their natural diffuse characteristics. Despite the superiority of
the ETOs and in particular of the CS functions, which constitute a
complete orthogonal set, the ETOs (as well as the GTOs) are by no
other means optimal for excited states description of many-electron
systems.

Hence, a common way of building custom basis sets for excited states
is to extend a standard GTO basis with a large set of primitive
diffuse functions. The primitives are often put in the form of the
even tempered Gaussians (ETG), in which the exponents are given by a
simple formula
\begin{equation}
	log \, \zeta_k = log \, \zeta_0 + k\,log \, \alpha;\;  log \, \alpha < 0,\,
        k = 0 \, \dots \, (N-1)
\end{equation}
where $\zeta_0$ and $\alpha$ are the ETG series parameters and $N$ is
the size of the set.  Such a series can provide many diffuse functions
with just a few variable parameters. However, the drawback of adding
extra functions to a basis is that they tend to cause near linear
dependencies, leading to serious numerical issues. A disadvantage of
the ETG sets, in particular, is that they do not consistently cover a
larger portion of the spectrum and are hardly applicable to problems
with excessive demands on precision and flexibility like in dynamical
studies of electronic resonances under complex scaled
representation.~\cite{13KaSm} By extending a basis with extra diffuse
functions it is practically sufficient that the resulting set is not
linearly dependent and satisfactorily describes a particular feature
of interest.

The aim of the present study is a tailored optimization of the diffuse
basis rather than a mere inclusion of a large number of functions. The
intention is to systematically generate such a basis that would
approximate at least a few Rydberg orbitals. Such orbitals could then
serve as the optimal functions for a correlated treatment of the
corresponding Rydberg states. The ultimate outcome, although beyond
the scope of this work, might lead to a specific diffuse basis
suitable for a complex scaling treatment of related resonance states,
which is highly sensitive to the basis set quality.~\cite{13KaSm} In
order to employ standard quantum chemistry codes this study
exclusively uses Gaussian basis sets. Nevertheless, the investigated
approach is universal and applicable to any basis set type.

This study focuses on systems that can be modeled as a closed-shell
ion core with an odd electron moving around. For such systems a simple
trick can be used to describe the Rydberg orbitals that can be the
target of the basis set optimization. As is well known from the
Koopmans' theorem,\cite{SzaboOstlund_1996} the canonical virtual
orbitals of a closed-shell system describe an electron captured by the
system ({\em i.e.} the Rydberg electron captured by the ion core).
For lithium atom (Li), which is the subject of this work, it means
that by a variational minimization of the Li$^+$ virtual orbital
energies, while varying the basis set parameters, one should end up
with such Li$^+$ virtual orbitals that could serve as the appropriate
basis functions for the description of the Rydberg excited electron in
the neutral Li atom.

Although the suggested approach is based on the Hartree--Fock model
only, it can be anticipated that due to the Rydberg character of the
excited electron, the model can be satisfactory at least for higher
excited states. In our earlier studies of helium (He)
resonances,\cite{13KaSm,13KaSmCi} we used an analogous approach to
obtain a basis for He excited states by appropriately modifying the
Fock operator so that the virtual orbitals describe excited electrons
rather than the captured ones. The quality of the optimized basis
was outstanding, leading not only to good transition energies, but
also to a wide interval of resonance energy stability along the
complex scaling parameter $\vartheta$. Such a basis enabled an
extensive non-Hermitian dynamical propagation of He under extremely
intense laser radiation.

The rest of the article is organized as follows.
Sect.~\ref{Computational} describes the basis set optimization
and other computational details. Sect.~\ref{Results} discusses
important properties of the optimized basis and compares the resulting
Rydberg excitation energies to other highly accurate values known in
the literature. Summary of the results and conclusions are drawn in
Sect.~\ref{Conclusions}.

\section{Computational Details}
\label{Computational}
The whole basis set optimization process can be summarized in the
following. First, a standard basis set is included for a proper
description of the electronic correlation. That should suffice for
the ground as well as for the Rydberg states since the part of
the wave function, in which the correlation plays role, is spatially
distributed similarly; only the Rydberg electron occupies distant
areas where it does not significantly contribute to the dynamical
electronic correlation.
To describe the Rydberg electron attracted by a closed-shell ion core,
diffuse functions are appended to the basis and the virtual orbital
energies of the ion core are minimized. This is achieved by varying
the diffuse functions while keeping the optimal Hartree--Fock energy
and avoiding near linear dependencies. In accord with
Koopmans' theorem, such optimal virtual orbitals are just the Rydberg
orbitals. Since their shape is controlled solely by the Hartree--Fock
field of the ion core, it is advantageous to improve the field quality
before the actual virtual orbital optimization by an additional high
exponent function to mimic the wave-function cusp. Eventually,
contracted basis functions are formed from the optimal Rydberg
orbitals using their expansion coefficients (LCAO) in the primitive
diffuse basis. The final basis set for productive computations of the
Rydberg states then consists of the standard basis together with the
high exponent function and appropriate subsets of the contracted
optimized Rydberg functions as discussed in Section~\ref{ResultsA}.

Practically, for lithium, the well established
aug-ano-pVQZ~\cite{11NeVa} basis of Neese and Valeev has been chosen
as the standard basis set. To improve the Hartree--Fock wave function
of the Li$^+$ ion core an additional ETG series of high exponent S
functions was used to minimize the Hartree--Fock energy. From the
resulting 1S orbital an auxiliary contracted function was formed,
consisting of the ETG primitives only, which was used further instead
of the high exponent series.  Independently for each angular momentum
$L \in \{{\rm S, P, D, F, G}\}$, a diffuse ETG set of the given $L$
was added to the basis and by varying its parameters the energy of the
first Li$^+$ virtual orbital of that $L$ symmetry was minimized. With
such optimized ETG parameters only the number ($N$) of the ETG
primitive functions was further gradually increased until a
satisfactorily large number of virtual orbital energies were no longer
changing by more than $10^{-9}$~a.u..  In the end, all the optimized
virtual orbitals, namely 25~S, 28~P, 22~D, 21~F and 14~G, were
contracted
in the diffuse ETG subspace
and put together in a final huge basis denoted as
$\Basis{25S-28P-22D-21F-14G}$. This notation specifies how many
optimal contracted Rydberg functions are added to the aug-ano-pVQZ
basis together with the high exponent contracted S function.

During the basis set optimization process the Hartree--Fock orbital
energies were calculated using the MRCC~\cite{mrcc_orig,mrcc_2020}
program package while the multidimensional minimization itself was
driven by the mdoptcli~\cite{mdoptcli} utility, which uses procedures
from the GNU Scientific Library (GSL)~\cite{GSL}. All the correlated
computations using the coupled cluster (CCSD) and equation of motion
coupled cluster (EOM-CCSD) methods employed the
GAMESS~2021~R1~\cite{gamess_2020} package, recompiled to allow large
number of primitive basis functions.

\section{Results and Discussion}
\label{Results}

\subsection{Properties of the optimized basis}
\label{ResultsA}
It was found that the $\Basis{7S-6P-5D}$ subset of the huge
$\Basis{25S-28P-22D-21F-14G}$ basis exhibits already converged CCSD
ionization potential (IP) with respect to the basis set size and
similarly also all the Li bound state excitation energies computed at
the EOM-CCSD level (up to the state 8S). This is apparent in
Table~\ref{TblLiCETG}, which shows that by adding more of the
optimized diffuse S, P or D functions (basis $\Basis{10S-9P-8D}$) or
by including F and G functions (basis $\Basis{7S-6P-5D-4F-3G}$) to the
$\Basis{7S-6P-5D}$ basis, the IP as well as the excitation energy
values were no longer affected. The $\Basis{7S-6P-5D}$ basis can thus
be considered as a minimal saturated set that can safely be extended
to achieve higher Li Rydberg states.

\begin{table}
  \caption{\label{TblLiCETG} Comparison of correlated ground and bound
    state excitation energies of neutral doublet Li resulting from three
    contracted basis sets. The ground state 2S energy (ROHF and CCSD)
    is given in a.u. while the excitation energies (EOM-CCSD) as well
    as the ionization potential (CCSD) are in eV.}
  \resizebox{!}{0.95\height}{
  \begin{tabular}{c|r|r|r}
    \hline \hline
    state & $\Basis{7S-6P-5D}$ & $\Basis{7S-6P-5D-4F-3G}$ & $\Basis{10S-9P-8D}$ \\
    \hline \hline
    2S ROHF & -7.432,726,827,6 & -7.432,726,827,6 & -7.432,726,828,0 \\
    2S CCSD & -7.474,434,973,0 & -7.474,434,975,9 & -7.474,435,349,2 \\
    IP CCSD & 5.3877 & 5.3877 & 5.3877 \\
    \hline \hline
    2P & 1.8472 & 1.8472 & 1.8472 \\
    3S & 3.3704 & 3.3704 & 3.3704 \\
    3P & 3.8317 & 3.8317 & 3.8317 \\
    3D & 3.8754 & 3.8753 & 3.8754 \\
    4S & 4.3378 & 4.3378 & 4.3378 \\
    4P & 4.5186 & 4.5186 & 4.5186 \\
    4D & 4.5373 & 4.5373 & 4.5373 \\
    4F &        & 4.5379 &        \\
    5S & 4.7456 & 4.7456 & 4.7456 \\
    5P & 4.8341 & 4.8341 & 4.8341 \\
    5D & 4.8437 & 4.8437 & 4.8437 \\
    5F &        & 4.8440 &        \\
    5G &        & 4.8441 &        \\
    \hline
  \end{tabular}}
\end{table}

It should be stressed that the obtained diffuse basis was optimized
for Li Rydberg states and not for Li$^{+}$, although it was the
virtual orbitals of the cation that determined the Li Rydberg
functions. Therefore, Li$^{+}$ excited states are not converged with
respect to the basis. That can be seen from Table~\ref{TblLiPlusCETG},
where the not yet converged Li$^{+}$ states are described by the very
same basis sets as the converged states of the neutral Li in
Table~\ref{TblLiCETG}.

\begin{table}
  \caption{\label{TblLiPlusCETG} Comparison of correlated ground and
    excitation energies of singlet Li$^+$ resulting from three
    contracted basis sets. The ground state 1S energy (RHF and CCSD)
    is given in a.u. while the excitation energies (EOM-CCSD) are in
    eV.}
  \resizebox{!}{0.97\height}{
  \begin{tabular}{c|r|r|r}
    \hline \hline
    state & $\Basis{7S-6P-5D}$ & $\Basis{7S-6P-5D-4F-3G}$ & $\Basis{10S-9P-8D}$ \\
    \hline \hline
    1S RHF  & -7.236,415,117,9 & -7.236,415,117,9 & -7.236,415,117,9 \\
    1S CCSD & -7.276,442,000,8 & -7.276,442,002,0 & -7.276,442,304,5 \\
    \hline \hline
    2S & 60.8448 & 60.8448 & 60.8446 \\
    2P & 62.2642 & 62.2642 & 62.2610 \\
    3S & 69.2199 & 69.2199 & 69.2194 \\
    3D & 69.7706 & 69.7706 & 69.7474 \\
    3P & 69.8285 & 69.8285 & 69.8204 \\
    4S & 72.0843 & 72.0843 & 72.0826 \\
    4D & 72.3562 & 72.3561 & 72.3458 \\
    4P & 72.3774 & 72.3774 & 72.3753 \\
    \hline
  \end{tabular}}
\end{table}

As anticipated above, to achieve higher Li Rydberg states of a
particular angular momentum $L$, just more of the optimized Rydberg
functions of the given $L$ can be added to the minimal
$\Basis{7S-6P-5D}$ basis (schematically as $\BasisKL{7S-6P-5D}{L}$),
since the resulting correlated states are no longer affected by the
functions of other $L$. Truly, Table~\ref{TblLiSPD+S} shows that by
gradually increasing the number of Rydberg S functions included to the
minimal basis, the correlated bound states remain intact, only new S
states emerge as the basis grows. This, as well as a strong dominance
of the appropriate R1 EOM-CCSD amplitudes in these states confirm
their Rydberg character and hence also suitability of the presented basis
set optimization scheme.  Analogous results were obtained also for P,
F and G states. Only the D states computed with
$\BasisKL{7S-6P-5D}{\rm D}$ basis exhibited unsaturated behavior, as
shown in Table~\ref{TblLiSPD+D}. Although the states below 7D
(regardless of their symmetry) were unaffected by the additional D
functions, none of the higher D states could achieve a converged
excitation energy. That could mean the D states higher than 6D may
still need more robust basis for better electronic correlation
description than similarly excited S, P, F or G states and thus the D
states may not yet have the true Rydberg nature.

Moreover, for all states with the principal quantum number $n$ larger
than 7, the optimized Rydberg functions are state selective. That
means, only a single specific Rydberg function needs to be added to
the $\Basis{7S-6P-5D}$ basis to achieve the appropriate correlated
Rydberg state, reducing the necessary basis set size
dramatically. Except for D states, again, where the differences
between the state selective basis and the $\Basis{7S-6P-14D}$ basis
reached even 0.1 eV, all the other $L$ states exhibited negligible
errors, from $3 \times 10^{-9}$~eV for state 14G, to $2 \times
10^{-6}$~eV for state 10S.

\begin{table}
  \caption{\label{TblLiSPD+S} Comparison of correlated ground and
    bound state excitation
    energies of a neutral doublet Li above the state 7S for three contracted
    $\BasisKLIn{7S-6P-5D}{L={\rm S}}$ basis sets. The ground state 2S energy
    (ROHF and CCSD) is given in a.u. while the excitation energies (EOM-CCSD)
    as well as the ionization potential (CCSD) are in eV.}
  \begin{tabular}{c|r|r|r}
    \hline \hline
    state & $\Basis{7S-6P-5D}$ & $\Basis{16S-6P-5D}$ & $\Basis{25S-6P-5D}$ \\
    \hline \hline
    2S ROHF & -7.432,726,827,6 & -7.432,726,828,0 & -7.432,726,828,0 \\
    2S CCSD & -7.474,434,973,0 & -7.474,434,980,6 & -7.474,434,980,5 \\
    IP CCSD & 5.3877 & 5.3877 & 5.3877 \\
    \hline \hline
    7P  & 5.1070 & 5.1070 & 5.1070 \\
    7D  & 5.1097 & 5.1097 & 5.1097 \\
    8S  &        & 5.1540 & 5.1540 \\
    9S  &        & 5.2116 & 5.2116 \\
    10S &        & 5.2610 & 5.2610 \\
    11S &        & 5.2990 & 5.2990 \\
    12S &        & 5.3264 & 5.3264 \\
    13S &        & 5.3457 & 5.3457 \\
    14S &        & 5.3592 & 5.3592 \\
    15S &        & 5.3684 & 5.3684 \\
    16S &        & 5.3747 & 5.3747 \\
    17S &        &        & 5.3790 \\
    \hline
  \end{tabular}
\end{table}

\begin{table}
  \caption{\label{TblLiSPD+D} Comparison of correlated ground and
    bound state excitation energies of a neutral doublet Li above the state 7S
    for three contracted $\BasisKLIn{7S-6P-5D}{L={\rm D}}$
    basis sets. The ground state 2S energy (ROHF and CCSD) is given in
    a.u. while the excitation energies (EOM-CCSD) as well as the
    ionization potential (CCSD) are in eV.}
  \begin{tabular}{c|r|r|r}
    \hline \hline
    state & $\Basis{7S-6P-5D}$ & $\Basis{7S-6P-10D}$ & $\Basis{7S-6P-14D}$ \\
    \hline \hline
    2S ROHF & -7.432,726,827,6 & -7.432,726,827,6 & -7.432,726,827,6 \\
    2S CCSD & -7.474,434,973,0 & -7.474,435,037,8 & -7.474,435,042,6 \\
    IP CCSD & 5.3877 & 5.3877 & 5.3877 \\
    \hline \hline
    7P  & 5.1070 & 5.1070 & 5.1070 \\
    7D  & 5.1097 & 5.1106 & 5.1088 \\
    8D  &        & 5.1745 & 5.1655 \\
    9D  &        & 5.2165 & 5.1750 \\
    10D &        & 5.2582 & 5.2159 \\
    11D &        & 5.2719 & 5.2618 \\
    12D &        & 5.2828 & 5.2733 \\
    13D &        &        & 5.3004 \\
    14D &        &        & 5.3559 \\
    \hline
  \end{tabular}
\end{table}

\subsection{Li bound excited states}
In this section the computed bound Li Rydberg states are presented and
compared with the best non-relativistic results known in the literature.
The most appropriate data come from the systematic studies of Li
$^2$S, $^2$P and $^2$D states that employ the explicitly correlated
Gaussians (ECG) and consider also the effect of the finite nuclear
mass as well as the leading relativistic and QED
corrections.\cite{19BrBuStAd,12BuAd,11ShBuAd} These state-of-the-art
studies provide highly accurate results for most Li excited states
that had ever been computed.  In this work the comparison is made only
to the non-relativistic ECG results, which practically reach the
estimates of the exact non-relativistic
energies.\cite{08PuPa,09SiHa,09BuKoStAd,12WaYaQiDr,99Ki} Another
comparison is made to the Hylleraas-CI (Hy-CI) study,\cite{13RuMaFr} which
provides considerably fewer $^2$S, $^2$P and $^2$D states.
Nevertheless, the same study also presents Full-CI energies in optimal
Slater-type-orbital (STO) basis that fully cover the principal quantum
number $n=7$, covering all the states from 2S to 7S and up to 7I. The
results for the individual angular momenta are compared in
Table~\ref{TblFinalSCmp} (S), \ref{TblFinalPCmp} (P),
\ref{TblFinalDCmp} (D), \ref{TblFinalFCmp} (F) and \ref{TblFinalGCmp}
(G). The state energies are presented in a.u., while the excitation
energies with respect to the ground 2S state are in eV rounded to
$10^{-4}$~eV, for convenience. When appropriate, values corresponding
to the estimates of the exact non-relativistic energies are labeled
with a related bibliographic reference.

Table~\ref{TblFinalSCmp} shows the $^2$S states computed with the
$\Basis{25S-6P-5D}$ basis.  We can see the ground state energy of the
present study is still more than 3 milihartree above the highly
precise computations. This is well understandable due to the Gaussian
character of the basis, lack of any explicit electronic correlation,
only the CCSD level of theory describing the three-electron system and
also due to the relatively small basis size.
Nevertheless, when we compare the excitation energies, we can see that
the present Gaussian basis results are consistently only a few meV off
the ECG values up to the state 8S. From the state 9S the differences
increase (even change the sign) and we might only estimate the error to
reach up to tens or even a hundred meV for the highest achieved state
24S. Such a sudden drop in accuracy may be put down to the higher D
orbital space insufficiency, as discussed with the
Table~\ref{TblLiSPD+D}, since the D functions contribute to the S
states correlation energy via double excitations.
The results are consistent within a few meV also with the Hy-CI and
with the STO Full-CI excitation energies except that the latter
exhibits an order of magnitude error jump for its highest 7S and 8S states
with respect to the precise computations, while the Gaussian basis
results of the present study remained consistent with the ECG values.

\begin{table}
  \caption{\label{TblFinalSCmp} Comparison of $^2$S EOM-CCSD excitation energies
    of Li bound states
    obtained from the $\Basis{25S-6P-5D}$ basis with the extensive
    non-relativistic ECG computations~\cite{19BrBuStAd},
    the Full-CI in an optimized STO basis~\cite{13RuMaFr}
    and the results of the Hylleraas-CI computations.~\cite{13RuMaFr}
    Energy levels ($E$) are given in a.u.
    while the excitation energies ($EE$) as well as their differences
    ($\Delta EE$) are in eV. $\Delta EE$ is defined as
    $EE^{\Basis{25S-6P-5D}} - EE^{\rm \text{ref.}}$.
    The reference 2S energy values in a.u. are put in parentheses for convenience.
    Values corresponding also to the exact non-relativistic estimate are marked
    with an asterisk and the bibliographic reference.}
  \resizebox{!}{0.96\height}{
  \begin{tabular}{c|r|r||r|r||r|r||r|r}
    \hline \hline
    \multirow{2}{*}{state} & \multicolumn{2}{c||}{$\Basis{25S-6P-5D}$} & \multicolumn{2}{c||}{ECG~\cite{19BrBuStAd}} & \multicolumn{2}{c||}{STO Full-CI~\cite{13RuMaFr}} & \multicolumn{2}{c}{Hy-CI~\cite{13RuMaFr}} \\ \cline{2-9}
    & $E$ & $EE$ & $EE$ & $\Delta EE$ & $EE$ & $\Delta EE$ & $EE$ & $\Delta EE$ \\
    \hline \hline
    2S  & -7.4744350 &         & (-7.4780603)$^{*,}$\cite{08PuPa} &          & (-7.477192) &      & (-7.478058969) &  \\
    3S  & -7.3505733 & 3.3704  & 3.3732$^{*,}$\cite{08PuPa}       & -0.0027  & 3.3727   & -0.0022 & 3.3733         & -0.0028 \\
    4S  & -7.3150230 & 4.3378  & 4.3410$^{*,}$\cite{09SiHa}       & -0.0032  & 4.3406   & -0.0027 & 4.3413         & -0.0035 \\
    5S  & -7.3000352 & 4.7456  & 4.7486$^{*,}$\cite{09SiHa}       & -0.0030  & 4.7486   & -0.0030 & 4.7496         & -0.0040 \\
    6S  & -7.2923000 & 4.9561  & 4.9579$^{*,}$\cite{09SiHa}       & -0.0018  & 4.9595   & -0.0033 & 4.9612         & -0.0050 \\
    7S  & -7.2878069 & 5.0784  & 5.0795$^{*,}$\cite{09SiHa}       & -0.0011  & 5.1047   & -0.0263 & 5.0878         & -0.0094 \\
    8S  & -7.2850298 & 5.1540  & 5.1563                          & -0.0023  & 5.2109   & -0.0569 & 5.1611         & -0.0071 \\
    9S  & -7.2829112 & 5.2116  & 5.2079                          &  0.0037  &          &         &                &  \\
    10S & -7.2810981 & 5.2610  & 5.2442                          &  0.0167  &          &         &                &  \\
    11S & -7.2797011 & 5.2990  & 5.2708                          &  0.0282  &          &         &                &  \\
    12S & -7.2786917 & 5.3264  & 5.2907                          &  0.0357  &          &         &                &  \\
    13S & -7.2779819 & 5.3457  & 5.3062                          &  0.0396  &          &         &                &  \\
    14S & -7.2774894 & 5.3592  &                                 &          &          &         &                &  \\
    15S & -7.2771501 & 5.3684  &                                 &          &          &         &                &  \\
    16S & -7.2769175 & 5.3747  &                                 &          &          &         &                &   \\
    17S & -7.2767584 & 5.3790  &                                 &          &          &         &                &  \\
    18S & -7.2766498 & 5.3820  &                                 &          &          &         &                &  \\
    19S & -7.2765758 & 5.3840  &                                 &          &          &         &                &  \\
    20S & -7.2765254 & 5.3854  &                                 &          &          &         &                &  \\
    21S & -7.2764911 & 5.3863  &                                 &          &          &         &                &  \\
    22S & -7.2764676 & 5.3870  &                                 &          &          &         &                &  \\
    23S & -7.2764516 & 5.3874  &                                 &          &          &         &                &  \\
    24S & -7.2764405 & 5.3877  &                                 &          &          &         &                &  \\
    \hline
  \end{tabular}}
\end{table}

In Table~\ref{TblFinalPCmp}, we can see that the $^2$P results in the
basis $\Basis{7S-24P-5D}$ are remarkably close to the ECG values
within meV accuracy while the STO Full-CI excitation energies are
shifted by an order of magnitude. It can also be noticed that the
present Gaussian excitation energies are all closer to the ECG results
and with a very consistent difference than the Hy-CI values.  From
this trend we might speculate that the accuracy of the highest
achieved 25P state could be in units or at maximum in tens of meV.

\begin{table}
  \caption{\label{TblFinalPCmp} Comparison of $^2$P EOM-CCSD excitation energies
    of Li bound states
    with respect to the ground 2$^2$S state
    obtained from the $\Basis{7S-24P-5D}$ basis with the extensive
    non-relativistic ECG computations~\cite{12BuAd},
    the Full-CI in an optimized STO basis~\cite{13RuMaFr}
    and the results of the Hylleraas-CI computations.~\cite{13RuMaFr}
    Energy levels ($E$) are given in a.u.
    while the excitation energies ($EE$) as well as their differences
    ($\Delta EE$) are in eV. $\Delta EE$ is defined as
    $EE^{\Basis{7S-24P-5D}} - EE^{\rm \text{ref.}}$.
    The reference 2S energy values in a.u. are put in parentheses for convenience.
    Values corresponding also to the exact non-relativistic estimate are marked
    with an asterisk and the bibliographic reference.}
  \resizebox{!}{0.89\height}{
  \begin{tabular}{c|r|r||r|r||r|r||r|r}
    \hline \hline
    \multirow{2}{*}{state} & \multicolumn{2}{c||}{$\Basis{7S-24P-5D}$} & \multicolumn{2}{c||}{ECG~\cite{12BuAd}} & \multicolumn{2}{c||}{STO Full-CI~\cite{13RuMaFr}} & \multicolumn{2}{c}{Hy-CI~\cite{13RuMaFr}} \\ \cline{2-9}
    & $E$ & $EE$ & $EE$ & $\Delta EE$ & $EE$ & $\Delta EE$ & $EE$ & $\Delta EE$ \\
    \hline \hline
    2S  & -7.4744358 &         & (-7.4780603)$^{*,}$\cite{09BuKoStAd,08PuPa} &  & (-7.477192) &         & (-7.478058969) &  \\
    2P  & -7.4065515 & 1.8472  & 1.8478$^{*,}$\cite{12WaYaQiDr}   &  -0.0005 & 1.8660         & -0.0187 & 1.8479         & -0.0007 \\
    3P  & -7.3336219 & 3.8317  & 3.8343                          &  -0.0026 & 3.8513         & -0.0196 & 3.8353         & -0.0036 \\
    4P  & -7.3083789 & 4.5186  & 4.5217                          &  -0.0031 & 4.5391         & -0.0205 & 4.5238         & -0.0052 \\
    5P  & -7.2967853 & 4.8341  & 4.8374                          &  -0.0033 & 4.8542         & -0.0201 & 4.8415         & -0.0074 \\
    6P  & -7.2905206 & 5.0046  & 5.0080                          &  -0.0034 & 5.0245         & -0.0199 & 5.0094         & -0.0048 \\
    7P  & -7.2867570 & 5.1070  & 5.1104                          &  -0.0034 & 5.1278         & -0.0208 & 5.1224         & -0.0154 \\
    8P  & -7.2843201 & 5.1733  & 5.1767                          &  -0.0034 &                &         &                &  \\
    9P  & -7.2826463 & 5.2188  & 5.2221                          &  -0.0033 &                &         &                &  \\
    10P & -7.2814198 & 5.2522  & 5.2545                          &  -0.0023 &                &         &                &  \\
    11P & -7.2804485 & 5.2787  &                                 &          &                &         &                &  \\
    12P & -7.2796394 & 5.3007  &                                 &          &                &         &                &  \\
    13P & -7.2789788 & 5.3186  &                                 &          &                &         &                &  \\
    14P & -7.2785302 & 5.3309  &                                 &          &                &         &                &  \\
    15P & -7.2782993 & 5.3371  &                                 &          &                &         &                &  \\
    16P & -7.2779292 & 5.3472  &                                 &          &                &         &                &  \\
    17P & -7.2775941 & 5.3563  &                                 &          &                &         &                &  \\
    18P & -7.2773332 & 5.3634  &                                 &          &                &         &                &  \\
    19P & -7.2771260 & 5.3691  &                                 &          &                &         &                &  \\
    20P & -7.2769625 & 5.3735  &                                 &          &                &         &                &  \\
    21P & -7.2768372 & 5.3769  &                                 &          &                &         &                &  \\
    22P & -7.2767405 & 5.3796  &                                 &          &                &         &                &  \\
    23P & -7.2766658 & 5.3816  &                                 &          &                &         &                &  \\
    24P & -7.2766082 & 5.3832  &                                 &          &                &         &                &  \\
    25P & -7.2765638 & 5.3844  &                                 &          &                &         &                &  \\
    \hline
  \end{tabular}}
\end{table}

Similarly in Table~\ref{TblFinalDCmp}, the $^2$D states obtained from
the $\Basis{7S-6P-14D}$ basis are closest to the ECG results with
consistent differences in meV, while the differences from the Hy-CI
values are slightly less regular. The STO Full-CI excitation energies
are again shifted by an order of magnitude.  The excitation to the
state 7D (the highest D state achieved by all three comparative
studies) exhibits sudden deviation from the three reference
results. This, again, matches with the observed difficulties of the
higher D states as in Table~\ref{TblLiSPD+D}.

\begin{table}
  \caption{\label{TblFinalDCmp} Comparison of $^2$D EOM-CCSD excitation energies
    of Li bound states
    with respect to the ground 2$^2$S state
    obtained from the $\Basis{7S-6P-14D}$ basis with the extensive
    non-relativistic ECG computations~\cite{11ShBuAd},
    the Full-CI in an optimized STO basis~\cite{13RuMaFr}
    and the results of the Hylleraas-CI computations.~\cite{13RuMaFr}
    Energy levels ($E$) are given in a.u.
    while the excitation energies ($EE$) as well as their differences
    ($\Delta EE$) are in eV. $\Delta EE$ is defined as
    $EE^{\Basis{7S-6P-14D}} - EE^{\rm \text{ref.}}$.
    The reference 2S energy values in a.u. are put in parentheses for convenience.
    Values corresponding also to the exact non-relativistic estimate are marked
    with an asterisk and the bibliographic reference.}
  \begin{tabular}{c|r|r||r|r||r|r||r|r}
    \hline \hline
    \multirow{2}{*}{state} & \multicolumn{2}{c||}{$\Basis{7S-6P-14D}$} & \multicolumn{2}{c||}{ECG~\cite{11ShBuAd}} & \multicolumn{2}{c||}{STO Full-CI~\cite{13RuMaFr}} & \multicolumn{2}{c}{Hy-CI~\cite{13RuMaFr}} \\ \cline{2-9}
    & $E$ & $EE$ & $EE$ & $\Delta EE$ & $EE$ & $\Delta EE$ & $EE$ & $\Delta EE$ \\
    \hline \hline
    2S  & -7.4744350 &         & (-7.4780603)$^{*,}$\cite{09BuKoStAd,08PuPa} &         & (-7.477192) &         & (-7.478058969) &  \\
    3D  & -7.3320162 & 3.8754  & 3.8786$^{*,}$\cite{12WaYaQiDr}              & -0.0032 & 3.8937      & -0.0183 & 3.8789         & -0.0035 \\
    4D  & -7.3076902 & 4.5373  & 4.5408                                     & -0.0034 & 4.5560      & -0.0187 & 4.5402         & -0.0028 \\
    5D  & -7.2964306 & 4.8437  & 4.8472                                     & -0.0035 & 4.8624      & -0.0187 & 4.8482         & -0.0045 \\
    6D  & -7.2903160 & 5.0101  & 5.0137                                     & -0.0036 & 5.0288      & -0.0187 & 5.0167         & -0.0066 \\
    7D  & -7.2866912 & 5.1088  & 5.1140                                     & -0.0053 & 5.1291      & -0.0204 & 5.1226         & -0.0138 \\
    8D  & -7.2846058 & 5.1655  &                                            &         &             &  &  &  \\
    9D  & -7.2842566 & 5.1750  &                                            &         &             &  &  &  \\
    10D & -7.2827529 & 5.2159  &                                            &         &             &  &  &  \\
    11D & -7.2810660 & 5.2618  &                                            &         &             &  &  &  \\
    12D & -7.2806438 & 5.2733  &                                            &         &             &  &  &  \\
    13D & -7.2796484 & 5.3004  &                                            &         &             &  &  &  \\
    14D & -7.2776090 & 5.3559  &                                            &         &             &  &  &  \\
    \hline
  \end{tabular}
\end{table}

As for the $^2$F states (Table~\ref{TblFinalFCmp}), computed in the
$\Basis{7S-6P-5D-13F}$ basis, there are not many studies to compare
with. The estimates of the exact non-relativistic values are taken
from a relatively old review article by King.~\cite{99Ki} The
differences between the estimates and the present results are within a
few meV, that is still on par with the precise ECG studies of
Adamowicz {\it et al.}\cite{19BrBuStAd,12BuAd,11ShBuAd} for the $^2$S,
$^2$P and $^2$D states. The STO Full-CI results also differ by only a
few meV, although with the opposite sign. The reason why the $^2$F
excitation energies are this close to the STO Full-CI results in
contrast to the $^2$S, $^2$P or $^2$D states may arise from the true
Rydberg character of the $^2$F states.

\begin{table}
  \caption{\label{TblFinalFCmp} Comparison of $^2$F EOM-CCSD excitation energies
    of Li bound states
    with respect to the ground 2$^2$S state
    obtained from the $\Basis{7S-6P-5D-13F}$ basis with estimate of the exact
    non-relativistic values~\cite{99Ki} and the Full-CI results
    in an optimized STO basis.~\cite{13RuMaFr}
    Energy levels ($E$) are given in a.u.
    while the excitation energies ($EE$) as well as their differences
    ($\Delta EE$) are in eV. $\Delta EE$ is defined as
    $EE^{\Basis{7S-6P-5D-13F}} - EE^{\rm \text{ref.}}$.
    The reference 2S energy values in a.u. are put in parentheses for convenience.
    Values corresponding to the exact non-relativistic estimate are marked
    with an asterisk and the bibliographic reference.}
  \begin{tabular}{c|r|r||r|r||r|r}
    \hline \hline
    \multirow{2}{*}{state} & \multicolumn{2}{c||}{$\Basis{7S-6P-5D-13F}$} & \multicolumn{2}{c||}{Exact non-rel. est.~\cite{99Ki}} & \multicolumn{2}{c}{STO Full-CI~\cite{13RuMaFr}} \\ \cline{2-7}
    & $E$ & $EE$ & $EE$ & $\Delta EE$ & $EE$ & $\Delta EE$ \\
    \hline \hline
    2S  & -7.4744350 &         & (-7.4780603)$^{*,}$\cite{08PuPa} &         & (-7.477192) &  \\
    4F  & -7.3076701 & 4.5379  & 4.5414$^{*,}$\cite{99Ki}         & -0.0035 &  4.5329     & 0.0050 \\
    5F  & -7.2964194 & 4.8440  & 4.8475$^{*,}$\cite{99Ki}         & -0.0035 &  4.8396     & 0.0045 \\
    6F  & -7.2903080 & 5.0103  &                                 &         &  5.0064     & 0.0040 \\
    7F  & -7.2866230 & 5.1106  &                                 &         &  5.1100     & 0.0006 \\
    8F  & -7.2842313 & 5.1757  &                                 &         &             &  \\
    9F  & -7.2825915 & 5.2203  &                                 &         &             &  \\
    10F & -7.2814185 & 5.2522  &                                 &         &             &  \\
    11F & -7.2805499 & 5.2759  &                                 &         &             &  \\
    12F & -7.2798861 & 5.2939  &                                 &         &             &  \\
    13F & -7.2793602 & 5.3082  &                                 &         &             &  \\
    14F & -7.2789311 & 5.3199  &                                 &         &             &  \\
    15F & -7.2785961 & 5.3290  &                                 &         &             &  \\
    16F & -7.2783444 & 5.3359  &                                 &         &             &  \\
    \hline
  \end{tabular}
\end{table}

The only reference data for the $^2$G states in
Table~\ref{TblFinalGCmp} that we can compare to are those from the STO
Full-CI computations. Similarly like for the $^2$F states, the
results differ in the order of meV, although there are only
three excited states available for comparison.

\begin{table}
  \caption{\label{TblFinalGCmp} Comparison of $^2$G EOM-CCSD excitation energies
    of Li bound states
    with respect to the ground 2$^2$S state
    obtained from the $\Basis{7S-6P-5D-12G}$ basis with the Full-CI results
    in an optimized STO basis.~\cite{13RuMaFr}
    Energy levels ($E$) are given in a.u.
    while the excitation energies ($EE$) as well as their differences
    ($\Delta EE$) are in eV. $\Delta EE$ is defined as
    $EE^{\Basis{7S-6P-5D-12G}} - EE^{\rm \text{ref.}}$.
    The reference 2S energy value in a.u. is put in parentheses for convenience.}
  \begin{tabular}{c|r|r||r|r}
    \hline \hline
    \multirow{2}{*}{state} & \multicolumn{2}{c||}{$\Basis{7S-6P-5D-12G}$} & \multicolumn{2}{c}{STO Full-CI~\cite{13RuMaFr}} \\ \cline{2-5}
    & $E$ & $EE$ & $EE$ & $\Delta EE$ \\
    \hline \hline
    2S  & -7.4744350 &         & (-7.477192) &  \\
    5G  & -7.2964303 & 4.8437  & 4.8371      & 0.0066 \\
    6G  & -7.2903074 & 5.0104  & 5.0041      & 0.0062 \\
    7G  & -7.2866226 & 5.1106  & 5.1045      & 0.0061 \\
    8G  & -7.2842310 & 5.1757  &             &  \\
    9G  & -7.2825914 & 5.2203  &             &  \\
    10G & -7.2814184 & 5.2522  &             &  \\
    11G & -7.2805500 & 5.2759  &             &  \\
    12G & -7.2798892 & 5.2938  &             &  \\
    13G & -7.2793766 & 5.3078  &             &  \\
    14G & -7.2789649 & 5.3190  &             &  \\
    15G & -7.2785953 & 5.3291  &             &  \\
    16G & -7.2782391 & 5.3388  &             &  \\
    \hline
  \end{tabular}
\end{table}

\section{Conclusions}
\label{Conclusions}
For an {\it ab initio} quantum chemical description of Li Rydberg
states an appropriate diffuse Gaussian basis was developed by a
variational optimization of virtual orbitals of the corresponding
Li$^+$ ion core. The approach is based on Koopmans' theorem, which
says that a virtual orbital of a closed shell system describes a
captured electron, which actually means a Rydberg state.  The
resulting basis consists of the standard aug-ano-pVQZ set for
sufficient description of the electronic correlation, an extra high
exponent contracted function to achieve a high quality Hartree--Fock
field and a set of optimal diffuse Rydberg functions.

At the EOM-CCSD level of theory a minimal subset of the optimized
basis could be found for which the ionization potential as well as the
excitation energies were converged with respect to the number of the
Rydberg functions used.  Higher Rydberg states could be effectively
achieved by a state selective inclusion of the corresponding Rydberg
function, dramatically reducing the demands on the basis set size.

Dozens of states at high accuracy could be achieved by the present
approach, namely up to the states 24S, 25P, 14D, 16F and 16G, that is
many more than by competitive {\it ab initio} methods. Compared to the
state-of-the-art ECG approach, the computed excitation energies
consistently differed mostly in the order of meV. Only the D states
above 6D could not achieve convergence, which might be due to their
supposedly weak Rydberg character. The results were also comparable to
Hylleraas-CI energies, however, the differences were less regular than
with the ECG values. Nevertheless, the presented excitation energies even surpassed
accuracy of Full-CI results computed in an optimal STO basis by an
order of magnitude.

Regardless of the excellent quality of the presented basis, a plenty
of room for improvement remains. The number of primitive functions is
still too large for practical use in standard quantum chemistry codes,
which typically impose various restrictions on the basis set size. The
optimization process could also be more sophisticated, involve
directly more virtual orbitals and the basis could use more flexible
parameterization than ETG, {\it e.g.} ExTG~\cite{13KaSm}. On the other
hand, not all chemical systems would require this high accuracy for
the excited states so the optimization criteria might appropriately
loosen.

Although the study was performed on a lithium atom, the approach is
universal for any chemical system with a closed-shell ion core,
including molecules. Similarly, it is not limited to Gaussian basis
sets, but can be applied to any basis set types. Once the presented
tailored basis set generation process becomes sufficiently tuned so
that it is feasible for common quantum chemistry codes, it could
promise an affordable highly accurate approach to {\it ab initio}
state-selective investigation of molecular Rydberg states and possibly
also of their related resonances.



\begin{acknowledgments}
  The work was financially supported by the Grant Agency of
the Czech Republic (Grant 20-21179S).
  Computational resources were supplied by the project
  "e-Infrastruktura CZ" (e-INFRA CZ LM2018140 ) supported by the
  Ministry of Education, Youth and Sports of the Czech Republic.
\end{acknowledgments}

\bibliography{licetg}

\end{document}